\documentclass[useAMS,usenatbib]{mn2e}
\usepackage{amssymb}
\usepackage{deluxetable}
\usepackage{url}

\newcommand{\e}{\epsilon}

\newcommand{\psim}{\lower.5ex\hbox{$\; \buildrel \propto \over\sim \;$}}
\newcommand{\lbar}{\lower.0ex\hbox{$\; \buildrel
{\lower0.0ex \hbox{-}} \over\lambda  \;$}}

\newcommand{\cm}{\mathrm{cm}}

\newcommand{\erg}{\mathrm{erg}}

\newcommand{\s}{\mathrm{s}}

\newcommand{\txs}{TXS\,0536$+$145}

\title[The high-z $\gamma$-ray flaring blazar TXS\,0536$+$135]{Exploring
  the multi-band emission of TXS\,0536$+$145: the most distant
  $\gamma$-ray flaring blazar} 
\author[M. Orienti, F. D'Ammando, M. Giroletti et al.]
  {M. Orienti$^{1}$\thanks{E-mail: orienti@ira.inaf.it},
F. D'Ammando$^{1,2}$, M. Giroletti$^{1}$, J. Finke$^{3}$,
M. Ajello$^{4}$,
\newauthor D. Dallacasa$^{1,2}$, T. Venturi$^{1}$\\
$^1$INAF -- Istituto di Radioastronomia, via Gobetti 101, I-40129, Bologna,
Italy \\
$^{2}$Dipartimento di Fisica e Astronomia, Universi\`a degli Studi di
Bologna, via Ranzani 1, I-40127 Bologna, Italy\\
$^{3}$US Naval Research Laboratory, Code 7653, Washington, DC, USA\\ 
$^{4}$Department of Physics and Astronomy, Clemson University,
Clemson, SC 29634, USA\\ 
}
\date{Received \today; accepted ?}

\pagerange{\pageref{firstpage}--\pageref{lastpage}} \pubyear{2002}

\def\LaTeX{L\kern-.36em\raise.3ex\hbox{a}\kern-.15em
    T\kern-.1667em\lower.7ex\hbox{E}\kern-.125emX}

\begin{document}

\label{firstpage}

\maketitle

\begin{abstract}
We report results of a multi-band monitoring campaign of the 
flat spectrum radio quasar TXS\,0536$+$145 at redshift 2.69. 
This source was
detected during a very high $\gamma$-ray activity state in 2012 March 
by the Large Area Telescope on board {\it Fermi}, becoming 
the $\gamma$-ray flaring blazar at the highest redshift detected so
far. At the peak of the flare the source reached an apparent
isotropic $\gamma$-ray luminosity of 6.6$\times$10$^{49}$ erg s$^{-1}$
which is comparable to the values achieved by the most luminous blazars. 
This activity triggered radio-to-X-rays monitoring observations by {\it
  Swift}, Very Long Baseline Array, European VLBI Network, 
and Medicina single-dish telescope. 
Significant variability was observed from radio to X-rays
supporting the identification of the $\gamma$-ray source with
TXS\,0536$+$145.  
Both the radio and $\gamma$-ray light
curves show a similar behaviour, with the $\gamma$-rays leading
the radio variability with a time lag of about 4-6 months. The luminosity
increase is associated with a flattening of the radio spectrum.
No new superluminal component associated with the flare was detected in 
high resolution parsec-scale radio images. During the flare the $\gamma$-ray
spectrum seems to deviate from a power law, showing a curvature
that was not present during the average activity state.
The $\gamma$-ray properties of
TXS\,0536$+$145 are consistent with those shown by the high-redshift 
$\gamma$-ray blazar population. 
\end{abstract}


\section{Introduction}

High-redshift blazars are among the most powerful objects in the
Universe. Although they represent a significant fraction ($\sim$22 per
cent) of the
extragalactic hard 
X-ray sources \citep[e.g.][]{majello12}, they are 
 not commonly detected in $\gamma$-rays. 
Only 35 out of 360 flat spectrum radio quasars (FSRQ) detected by the
Large Area Telescope (LAT) on board {\it Fermi} and listed in the Second
LAT Catalogue of Active Galactic Nuclei (2LAC) have redshift $z >
2$ \citep{ackermann11}.
In the first LAT catalogue of $\gamma$-ray sources
above 10 GeV (1FHL) only 7 sources with $z >2$ are detected
\citep{ackermann13}. 
They are FSRQ already listed in the 2LAC sample, and represent the 20
per cent of the FSRQ with $z > 2$ present in the 2LAC.\\ 
Despite their low fraction in high-energy catalogues, the $\gamma$-ray
emission from high-redshift blazars is a powerful tool for probing 
the extragalactic background light (EBL). This diffuse radiation
consists of
photons likely produced by stars and galaxies across the Universe and
its lifetime. 
As the high-energy photons from extragalactic sources
propagate through the Universe, they interact with EBL photons at
ultraviolet through optical wavelengths by $\gamma$-$\gamma$
absorption, and create electron-positron pairs
\citep[e.g.][]{gould67}. This results in an attenuation of the
$\gamma$-ray sources above a critical energy that depends on the
redshift and the EBL model assumed \citep[e.g.][]{franceschini08}. \\
So far, no absorption feature related to the EBL has been
unambiguously observed in the spectra of 
individual sources. The EBL
attenuation has been also investigated by combining the spectra of a sample
of $\gamma$-ray BL Lacs detected above 3 GeV and spanning a redshift
range between 0.03 and 1.6 \citep{ackermann12}. For local sources ($z <
0.2$), no EBL attenuation is
observed up to 120 GeV. The absorption feature is
clearly visible for sources with $z > 0.2$, and evolves with
redshift. The lack of BL Lacs above $z > 1.6$ precluded
a statistical approach for investigating the EBL attenuation at higher
redshifts. This information can be obtained by the analysis of the
spectra of FSRQ, whose high redshift distribution extends beyond $z =
2.5$. The analysis of the FSRQ detected above 10 GeV suggested that
sources become softer with increasing redshift. This is particularly
evident for the FSRQ from the 1FHL with $z > 2$. The spectral softening
above 10 GeV (and not above 100 MeV) is interpreted as evidence of EBL
attenuation, although a cosmological evolution of the photon index
cannot be ruled out \citep{ackermann12}. \\
The detection of FSRQ at $z > 2$ during $\gamma$-ray flaring activity
may provide 
a step forward in our understanding of the EBL attenuation. During
$\gamma$-ray flaring episodes the spectra of FSRQ show a moderate
hardening \citep{abdo10c}, allowing us to explore energies that are
usually strongly attenuated due to the intrinsic source spectrum.
So far, 10 blazars at $z >
2$ have been detected during $\gamma$-ray flaring activity. 
One of these objects is
TXS\,0536$+$145, which was detected during a $\gamma$-ray flare
on 2012 March 22 by {\it Fermi}-LAT \citep{mo12}. 
With a redshift of $z=2.69$ \citep{sowards05}, this source is
the $\gamma$-ray flaring blazar at the highest
redshift observed so far.\\
TXS\,0536$+$145 was not part of the
1FGL or 2FGL catalogues \citep{abdo10a, nolan12}, 
indicating its low activity state during
the first two years of {\it Fermi}-LAT observations.
The high-energy flare triggered a multi-wavelength monitoring
campaign in X-rays, optical ({\it Swift}, XRT and UVOT) and radio 
bands (Very Long Baseline Array (VLBA), European VLBI Network (EVN) 
interferometers, and Medicina single-dish
telescope). In this paper we present the results of the multi-band
campaign. In particular we investigate the possible connection between
the $\gamma$-ray flare and flux variability in low-energy bands, as
well as changes in the radio jet structure. The information is then
used to determine the physical properties of this extreme object and
to model its spectral energy distribution (SED) during the flaring state.\\
The paper is organized as follows. In Section 2
we present the {\it Fermi}-LAT data
analysis and results. In Section 3 we report the radio data
collected by VLBA, EVN, and the Medicina single-dish telescope, 
while in Section 4 we present
the {\it Swift} observations performed in 2012 April and November. 
Results of the
multi-wavelength campaign are presented in Section 5. The
discussion and the presentation of the SED modelling is given in
Section 6, while a brief summary is in
Section 7. 

Throughout this paper, we assume the following cosmology: $H_{0} =
71\; {\rm km \; s^{-1} \; Mpc^{-1}}$, $\Omega_{\rm M} = 0.27$ and
$\Omega_{\rm \Lambda} = 0.73$, in a flat Universe. 
At the redshift of the target, $z$ = 2.69 \citep{sowards05}, 
the luminosity distance $D_{\rm L}$ is
22600 Mpc, and 1 milliarcsecond = 8.06
pc. \\

\section{{\it Fermi}-LAT Data: Selection and Analysis}
\label{Fermi}

The {\em Fermi}-LAT  is a pair-conversion telescope operating from 20 MeV to
$>$ 300 GeV. It has a large peak effective area ($\sim$8000 cm$^{2}$ for 1 GeV
photons), an energy resolution of typically $\sim$10 per cent, and a
field of view of 
about 2.4 sr with single-photon angular resolution (radius of 68 per cent 
containment angle) of
0\fdg6 at $E$ = 1 GeV on-axis. Details of the {\em
  Fermi}-LAT are given in \citet{atwood09}. \\ 
\begin{figure}
\begin{center}
\includegraphics{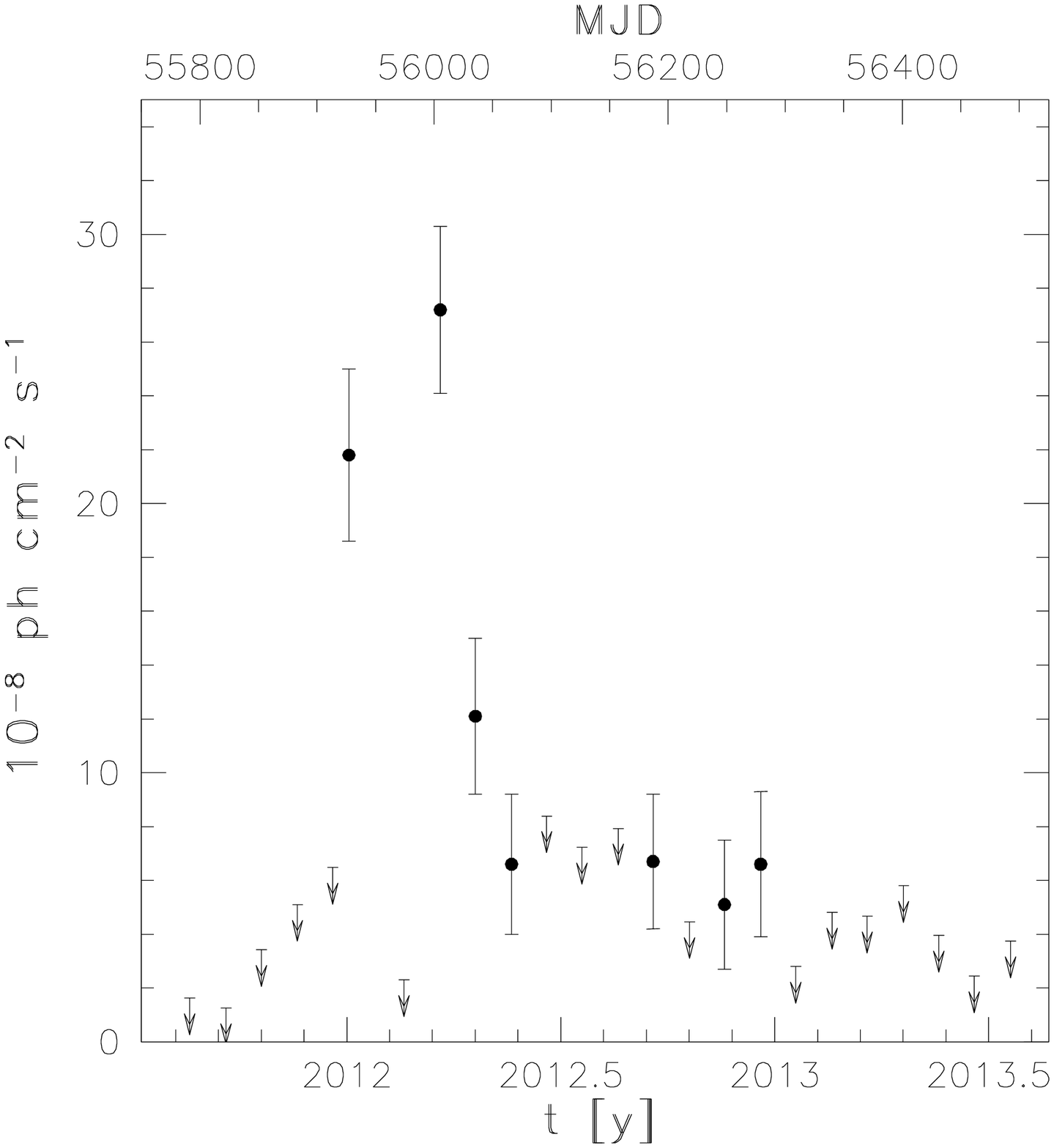}
\includegraphics{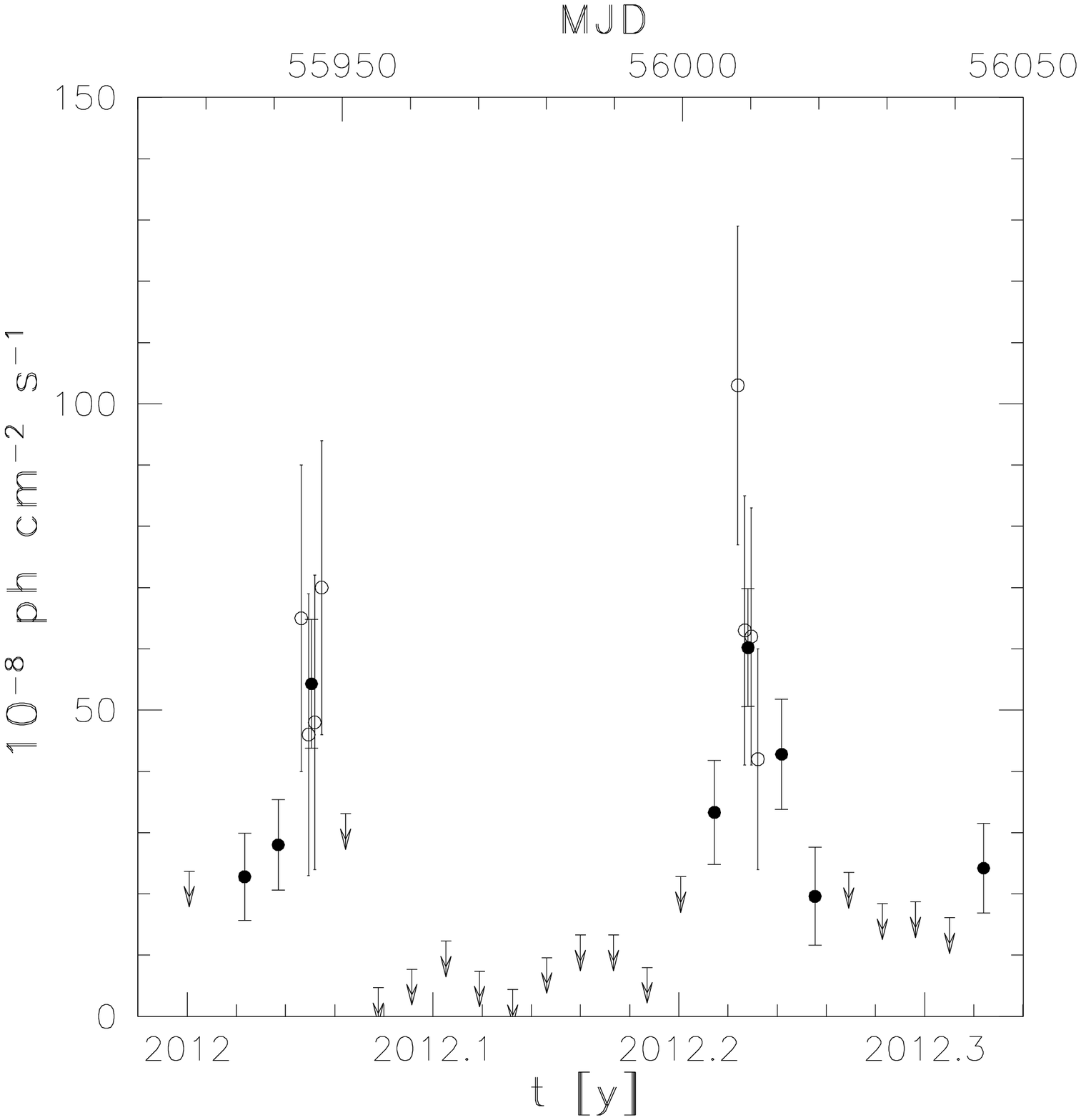}
\vspace{14cm}
\caption{{\it Top}: Integrated LAT light curve of TXS\,0536$+$145
  obtained in the 
  0.1--100 GeV energy range during 2011 August 4 -- 2013 August 4 with
  1-month time bins. Arrows refer to 2-$\sigma$ upper limits on the
  source flux. Upper limits are computed when TS $<$ 10. {\it Bottom}: Integrated LAT light curve of TXS\,0536$+$145 obtained in the
  0.1--100 GeV 
energy range during 2012 January 1 -- 2012 April 30 with 5-day time
bins. Open circles refer to daily flux for the high activity periods.
Arrows refer 
to 2-$\sigma$ upper limits on the source flux. Upper limits are
computed when TS $<$ 
10.}
\label{LAT}
\end{center}
\end{figure}
The LAT data reported in this paper were collected from 2008 August 4 (MJD
54682) to 2013 August 4 (MJD 56508). During this time the LAT instrument
operated almost entirely in survey mode. The analysis was performed with the
\texttt{ScienceTools}\footnote{\url{http://fermi.gsfc.nasa.gov/ssc/data/analysis/}} software package version v9r32p5. 
Only events belonging to the ``Source''
class were selected. In addition, a cut on the
zenith angle ($< 100^{\circ}$) was applied to reduce contamination from
the Earth limb $\gamma$-rays, which are produced by cosmic rays interacting
with the upper atmosphere. Moreover, only photons detected when the spacecraft
rocking
angle was $<$52$^{\circ}$ were selected. The spectral analysis was
performed with the 
instrument response functions
\texttt{P7REP\_SOURCE\_V15} 
using the unbinned maximum likelihood method implemented
in the tool \texttt{gtlike}. A Galactic diffuse emission model
and the isotropic 
component, which is the sum of
an extragalactic and residual cosmic-ray background, were used to model the
background\footnote{\url{http://fermi.gsfc.nasa.gov/ssc/data/access/lat/BackgroundModels.html}}. The
normalizations of both components in the background model were
allowed to vary 
freely during the spectral fitting. \\
We analysed a region of interest of $10^{\circ}$ radius centred at the
location of TXS\,0536$+$145. We evaluated the significance of the $\gamma$-ray signal from the sources by
means of the maximum likelihood Test Statistic TS =
2$\Delta$log(likelihood) between models with
and without a point source at the position of TXS\,0536$+$145
\citep{mattox96}. The source model used in
\texttt{gtlike} includes all the point sources from the 2FGL catalogue
and from the preliminary list of the Third Fermi LAT Catalogue
(Ackermann et al., in preparation) that
fall within $15^{\circ}$ from TXS\,0536$+$145. The spectra of these sources
were parametrized by power-law functions, $dN/dE \propto$
$(E/E_{0})^{-\Gamma_{\gamma}}$, or log-parabola, $dN/dE \propto$
$E/E_{0}^{-\alpha-\beta \, \log(E/E_0)}$, as
reported in the 2FGL catalogue for the different sources \citep{nolan12}. In the
same way a power law with
super-exponential cut-off has been used for the bright pulsars in the field.
A first maximum likelihood analysis was performed to remove from the model the
sources having
TS $<$ 10 and/or a predicted number of counts based on the fitted model
$N_{pred} < 3 $. A second maximum likelihood analysis was performed
on the updated source model. The fitting procedure was performed with the
sources within 10$^{\circ}$ from
TXS\,0536$+$145 included with the
normalization factors and the photon indices left as free parameters. For the
sources located between 10$^{\circ}$ and 15$^{\circ}$ from our target we kept the
normalization and the photon index fixed to the values of the 2FGL catalogue.\\
Integrating over the first two years of
{\em Fermi} operation the fit yielded a TS = 2, with a 2-$\sigma$ upper limit
of 1.0$\times$10$^{-8}$ ph cm$^{-2}$ s$^{-1}$ in the 0.1--100 GeV
energy range and assuming $\Gamma_{\gamma}=2.37$. In 
the same way
integrating in the period 2010 August 4 -- 2011 August 4 (MJD 55412--55777)
the fit yielded a TS = 2, with a 2-$\sigma$ upper limit of
4.5$\times$10$^{-9}$ ph cm$^{-2}$ s$^{-1}$ in the 0.1--100 GeV energy
range. By contrast, the fit with a power-law model
to the data integrated over the fourth and fifth years of {\em Fermi} operation
(2011 August 4 -- 2013 August 4; MJD 55777--56508) in the 0.1--100 GeV
energy range results in a TS = 142, with an integrated average flux of
(4.2 $\pm$ 0.6) $\times$10$^{-8}$ ph cm$^{-2}$ s$^{-1}$ and a
photon index $\Gamma_{\gamma}$ = 2.37 $\pm$ 0.09. 
The upper panel of Fig. \ref{LAT} shows the $\gamma$-ray light curve
for the period 2011 
August 4 --2013 August 4 using a power-law model and 1-month time bins. For each
time bin the spectral parameters
of TXS\,0536$+$145 and all sources within 10$^{\circ}$ from the target 
were frozen to the
value resulting
from the likelihood analysis over the entire period. If TS $<$ 10 the value of
the fluxes were replaced by the 2-$\sigma$ upper limits. \\
On a monthly time scale the source was detected for the
first time in 2012 January, with an increasing activity in 2012 March followed
by a lower flux in 2012 April-May. After this period the source was
detected at a low flux level 
only in 2012 September, November, and December.\\
Leaving the photon index free to vary during the month
of the highest
activity (2012 March 4 -- April 4) the fit results in a photon index
$\Gamma_{\gamma}$ = 2.05 $\pm$ 0.08 and a TS=183. \\
We produced a light curve focused on the
period of high activity (2012 January - April) with 
5-day and 1-day time bins (Fig. \ref{LAT}, bottom panel). 
We used a 1-day time bin
for the period 
with higher statistics. 
A preliminary measurement of the peak flux on 22 March was reported by
\citet{mo12}. The daily averaged flux in the detailed analysis was
(1.0 $\pm$ 0.3)$\times$10$^{-6}$ ph cm$^{-2}$ s$^{-1}$, a factor of
$\sim$25 higher than 
the average flux estimated considering together 
the fourth and fifth years of {\em Fermi}-LAT
observations. The corresponding observed apparent isotropic $\gamma$-ray
luminosity peak is 
6.6$\times$10$^{49}$ erg s$^{-1}$ in the 0.1--100 GeV energy range, comparable to
the values reached by the most powerful FSRQ during flaring activity.\\
The $\gamma$-ray point source localization was determined by the use
of the \texttt{gtfindsrc}
tool applied to the $\gamma$ rays extracted over the period 2011
August 4 -- 2013 
August 4, and results in R.A. = 84\fdg908, Dec. = 14\fdg604
(J2000), with a 95\% error circle radius of 0\fdg07, at an angular
separation of 0\fdg05 from the radio position of TXS\,0536$+$145. This
likely indicates a physical association between the $\gamma$-ray source and the
low-energy counterpart. \\
In order to investigate the presence of a curvature in the
$\gamma$-ray spectrum 
of TXS\,0536$+$145 an alternative spectral model to the power law (PL), a log
parabola (LP) was used for the fit. By analysing the fourth and fifth
years of {\it Fermi}-LAT observations we obtained a spectral slope
$\alpha$ = 2.05 $\pm$ 0.09 at the reference energy $E_0$ = 300 MeV, a
curvature parameter around the peak 
$\beta$ = 0.11$\pm$0.07, with a TS = 142 and an integrated average flux of (3.2
$\pm$ 0.5)$\times$10$^{-8}$ ph cm$^{-2}$ s$^{-1}$ in the 0.1 -- 100
GeV energy range. 
We used a likelihood ratio test 
to check the PL model (null hypothesis) against the LP model (alternative
hypothesis). These values can be compared by defining the curvature
Test Statistic TS$_{\rm curve}$=TS$_{\rm LP}$--TS$_{\rm
  PL}$, which we find to be TS$_{\rm curve}$=0. Therefore,
no statistical evidence of a curved spectral shape is detected.\\
If we consider only the flaring period (2012 March), using a
log-parabola model we obtain a spectral slope
$\alpha$ = 1.52 $\pm$ 0.13 at the reference energy $E_0$ = 300 MeV, a
curvature parameter around the peak 
$\beta$ = 0.16$\pm$0.04, with a TS = 202.
In this case we have
TS$_{\rm curve}$=TS$_{\rm LP}$--TS$_{\rm PL}$= 19 corresponding to
$\sim$4.4-$\sigma$, indicating statistical evidence of a
curved spectral shape. 
By means of the \texttt{gtsrcprob} tool we estimated that the
highest energy 
photon emitted by TXS\,0536$+$145 
(with probability $>$ 80 per cent of being associated with the source)
 was observed on 
2012 March 22 at a distance of 0\fdg05 from the source with an energy of
11.2 GeV. The highest energy photon was detected at the peak of the
$\gamma$-ray high activity state.\\
Analysing the {\it Fermi}-LAT data with $E >$10 GeV 
collected between 2011 August and 2013 August, the fit with a power-law
model resulted in a TS=5, indicating that the source was not
detected at such high energy. We
evaluated the 2-$\sigma$ upper limit as
9.3$\times$10$^{-11}$ ph cm$^{-2}$ s$^{-1}$ (assuming $\Gamma_{\gamma}
=2.37$).\\

\section{Radio data}

\subsection{VLBA observations}
\label{vlba_obs}

Multi-frequency Target of Opportunity (ToO) VLBA observations (project code
BO042) of TXS\,0536$+$145 triggered by the $\gamma$-ray flare were
carried out at 
8.4, 15, and 24 GHz during five observing epochs between 2012 May and
2013 January, with a recording bandwidth of 16 MHz at 512
Mbps data rate. 
During each observing epoch, the source was observed for 20 min at 8.4
GHz, and for 40 min at 15 and 24
GHz, spread into several scans of about 4 min each, and cycling
through frequencies in order to improve the {\it 
  uv}-coverage. For this reason the flux density measurements of the
pc-scale emission at the
various frequencies can be considered roughly simultaneous during each epoch. 
The observing
epochs are separated by about 2 months.\\
Data reduction was performed using the NRAO's Astronomical Image
Processing System (\texttt{AIPS}). 
After the application of system temperature and antenna gain
information, the amplitudes were checked using the data on DA\,193
(J0555+398) which is unresolved on a large subset of baselines at all
frequencies, and whose flux density is monitored at the
JVLA\footnote{\url{http://www.aoc.nrao.edu/~smyers/evlapolcal/polcal_master.html}}. 
DA\,193 was also used to generate the bandpass correction. The source
PKS\,0528$+$134 was used as phase calibrator.\\
The uncertainties on the amplitude calibration were found 
to be approximately 
7 per cent at 8.4 and 15 GHz, and about 10 per cent at 24 GHz.
The target source is
strong enough to allow the fringe fitting with a solution interval of
one minute to preserve the phase coherence. Final images were produced
after a number of phase self-calibration iterations. Amplitude
self-calibration was applied using a solution interval longer than the
scan length to remove residual systematic errors at the end of the
self-calibration process. The 1-$\sigma$ noise
level measured on the image plane is about 0.15 - 0.17 mJy/beam, being
worse at higher frequency.
For a source as bright as our target, the
error on the flux density is dominated by the calibration uncertainty.\\
The restoring beam is about
2.2$\times$0.8 mas$^2$, 1.4$\times$0.5 mas$^2$, and 0.8$\times$0.2
mas$^2$ at 8.4, 15, and 24 GHz, respectively. \\
At 15 GHz we complemented our VLBA data with five additional
observations from the Monitoring Of Jets in Active galactic nuclei
with VLBA Experiments (MOJAVE) programme\footnote{The MOJAVE data
   archive is maintained at \url{http://www.physics.purdue.edu/MOJAVE}.} 
\citep{lister09} performed between 2012 April and 2013
August. Uncertainties on the flux density scale are approximately 5 per cent
\citep{lister13}. 
Results on the ToO VLBA and MOJAVE observations are reported in 
Table \ref{fluxvlba}.\\ 
In addition to the ToO observations, we retrieved four epochs of archival VLBA
data obtained in 2011 at 8.4 GHz (project code BC196) to study the light curve
before the $\gamma$-ray flare (see Table \ref{histo}). The calibration
and data reduction were performed in the same way as described
above. In these observing runs, TXS\,0536$+$145 was observed for a few
minutes as a phase calibrator, precluding an adequate {\it
  uv}-coverage for investigating changes in the radio structure,
whereas the data turned out to be useful to measure the pc-scale flux density.\\

\begin{table}
\caption{VLBA flux density of TXS\,0536$+$145. Column 1: observing date;
  Cols. 2, 3, and 4: VLBA flux density at 8.4, 15, and 24 GHz,
  respectively; Col. 5: spectral index between 8.4 and 15 GHz; Col. 6:
spectral index between 15 and 24 GHz.}
\begin{center} 
\begin{tabular}{cccccc}
\hline
Obs. date&$S_{8.4}$&$S_{15}$&$S_{24}$&$\alpha_{8.4}^{15}$&$\alpha_{15}^{24}$\\
         &  mJy   &  mJy   &  mJy  &                  &  \\
\hline
29/04/2012 & -   & 777$\pm$39\tablenotemark{a} & - & - & -\\
22/05/2012 & 479$\pm$33 & 830$\pm$58 & 926$\pm$93 & -0.95$\pm$0.17 & -0.25$\pm$0.25 \\
23/07/2012 & 576$\pm$40 & 661$\pm$46 & 705$\pm$70 & -0.25$\pm$0.17 & -0.15$\pm$0.25 \\
26/09/2012 & 686$\pm$48 & 804$\pm$56 & 927$\pm$93 & -0.30$\pm$0.17 & -0.30$\pm$0.25 \\
19/11/2012 & 758$\pm$53 & 833$\pm$58 & 785$\pm$78 & -0.20$\pm$0.17 &  0.10$\pm$0.25\\
21/11/2012 &  -  & 805$\pm$40\tablenotemark{a} &  -  &   -   &  -  \\
13/01/2013 & 761$\pm$53 & 719$\pm$50 & 708$\pm$70 &  0.10$\pm$0.17 &  0.05$\pm$0.25\\ 
10/02/2013 &  -  & 626$\pm$31\tablenotemark{a} &  -  &   -   &  -  \\
02/06/2013 &  -  & 498$\pm$25\tablenotemark{a} &  -  &   -   &  -  \\
12/08/2013 &   -  & 488$\pm$25\tablenotemark{a} &  -  &   -   &  -  \\
\hline
\end{tabular}
\end{center}
\tablenotetext{a}{MOJAVE data}
\label{fluxvlba}
\end{table}

\begin{table}
\caption{Archival VLBA data at 8.4 GHz of TXS\,0536$+$145. Column 1:
  observing date; Col. 2: on-source observing time; Col. 3: flux
  density at 8.4 GHz.}
\begin{center}
\begin{tabular}{cccc}
\hline
Obs. date&Obs. time & $S_{8.4}$\\
         &  min     & mJy \\
\hline
01/02/2011& 2.3& 275$\pm$14 \\
30/05/2011& 3.2& 260$\pm$13 \\
06/08/2011& 4.7& 280$\pm$14 \\
01/09/2011& 3.2& 200$\pm$10 \\
\hline
\end{tabular}
\end{center}
\label{histo}
\end{table}

\begin{table*}
\caption{EVN observations of TXS\,0536$+$145. Column 1: observing date;
  Col. 2: 22 GHz flux density; Cols. 3 and 4: beam size and beam
  position angle, respectively; Col. 5: 1-$\sigma$ noise level
  measured on the image plane; Col. 6: Antennas. Ef=Effelsberg;
  On=Onsala; Yb=Yebes; Sv=Svetloe; Zc=Zelenchukskaya; Ur=Urumqi;
  Hh=Hartebeesthoek; Nt=Noto; Sh=Shangai; Md=Medicina; Tr=Torun;
  Mh=Mets\"ahovi; Bd=Badary.}
\begin{center}
\begin{tabular}{cccccc}
\hline
Obs. date&$S_{22}$& Beam size& PA & rms &Notes\\

         &  mJy   &  mas$^{2}$   &  deg  & mJy/beam&                 \\
\hline
06/06/2012 & 646$\pm$64 & 0.65$\times$0.23& 0.11&0.5&Ef,On,Yb,Sv,Zc,Ur,Hh  \\
06/11/2012 & 721$\pm$72 & 1.41$\times$0.17& 5.62&0.3&Ef,On,Yb,Nt,Zc,Ur,Sh  \\
27/02/2013 & 530$\pm$53 & 1.21$\times$0.23& 4.11&0.2&Ef,On,Md,Nt,Tr,Yb,Mh,Ur,Sh\\
11/06/2013 & 490$\pm49$ & 0.56$\times$0.47&59.17&0.3&Ef,On,Md,Nt,Tr,Yb,Zc,Hh\\
17/10/2013 & 421$\pm$42 & 0.53$\times$0.22&-2.64&0.4&Ef,On,Md,Nt,Tr,Yb,Sv,Zc,Bd,Ur,Sh,Hh\\
\hline
\end{tabular}
\end{center}
\label{fluxevn}
\end{table*}

\subsection{EVN observations}
\label{evn_obs}

\begin{figure}
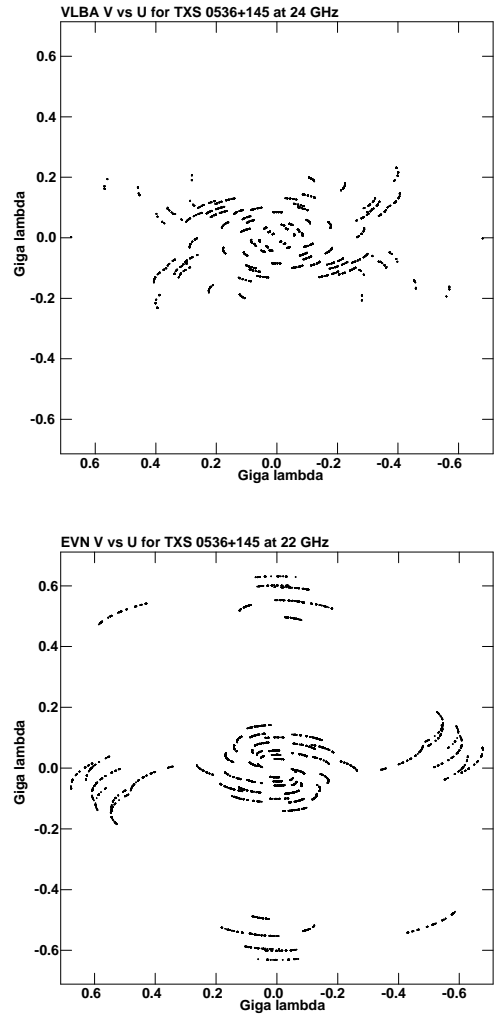

\begin{center}
\includegraphics{0536K_VLBAUV.PS}
\includegraphics{0536JUN13_UV.PS}
\vspace{14cm}
\caption{{\it Top}: VLBA $uv$-coverage at 24 GHz; {\it Bottom}: EVN
  $uv$-coverage at 22 GHz. The plots represent the $uv$-coverage
  of the VLBI network in units of wavelengths.}
\label{uvcoverage}
\end{center}
\end{figure}

Target of Opportunity EVN observations at 22 GHz were triggered by the
$\gamma$-ray flare and carried out between 2012 June and 2013
October, during the standard EVN sessions.
The observations were performed using
eight sub-bands separated by 16 MHz each for an aggregate bit rate of
1 Gbps. The telescopes involved in the
observing runs that provided good data are listed in Table
\ref{fluxevn}. Fig. \ref{uvcoverage} displays the VLBA and EVN
$uv$-coverages. It is worth noting the lack of baselines shorter than
40 M$\lambda$ in the EVN $uv$-coverage that prevents the detection of
features larger than $\sim$5 mas. On the other hand, the lack of
baselines longer than 200 M$\lambda$
in the North-South direction of the VLBA $uv$-coverage decreases the
resolution along that direction due to the elongated shape of the
observing beam.\\
Data reduction was performed using \texttt{AIPS}. A priori amplitude
calibration was derived with the \texttt{AIPS} task APCAL on the basis of the
measurements of the system temperatures and the antenna gain
information for each antenna. The flux density scale was checked by
means of the calibrator PKS\,0528$+$134 which was observed as a
calibrator also with the VLBA. The uncertainty on the flux density 
is approximately 10 per cent.
PKS\,0528$+$134 was also used as bandpass calibrator. \\
TXS\,0536$+$145 is strong enough to allow the fringe fitting with a
solution interval of 1 min to preserve the phase coherence. Final
images were produced after a number of phase self-calibration
iterations. \\

\subsection{Medicina observations}

After the 2012 April $\gamma$-ray flare, TXS\,0536$+$145 was observed
five times
with the Medicina single-dish telescope at 5 and 8.4 GHz.
Observations were performed with the new Enhanced Single-dish
Control System, which provides enhanced sensitivity and supports
observations with the cross scan technique. At each frequency the
typical on source time is 40 seconds and the flux density was
calibrated with respect to 3C\,286, 3C\,48, and NGC\,7027. 
Since the signal-to-noise ratio in each scan across the source was low
(typically ∼3), we performed a stacking analysis of the scans, which
allowed us to significantly improve the signal-to-noise ratio and the
accuracy of the measurement. 
The flux densities at 5 and 8.4 GHz
measured with the Medicina telescope are listed in Table \ref{medicina}. \\

\begin{table}
\caption{Medicina observations of TXS\,0536$+$145.}
\begin{center}
\begin{tabular}{ccc}
\hline
Date& $S_{\rm 5 GHz}$&$S_{\rm 8.4 GHz}$\\
    &   mJy       &  mJy \\
\hline
20/04/2012& 275$\pm$20& 365$\pm$15\\
21/05/2012&  - & 440$\pm$30\\
02/08/2012&  - & 470$\pm$40\\
06/09/2012&  - & 600$\pm$40\\
05/10/2012& 550$\pm$10&750$\pm$50\\
\hline
\end{tabular}
\end{center}
\label{medicina}
\end{table}

\begin{table*}
\caption{Log and fitting results of {\em Swift}/XRT observations of
  TXS\,0536$+$145 
using a power-law model with $N_{\rm H}$ fixed to Galactic
  absorption.}
\begin{center}
\begin{tabular}{ccccc}
\hline 
\multicolumn{1}{c}{Obs. date} &
\multicolumn{1}{c}{Net Exposure Time} &
\multicolumn{1}{c}{Net Count Rate} &
\multicolumn{1}{c}{Photon index} &
\multicolumn{1}{c}{Flux 0.3--10 keV\tablenotemark{a}} \\
\multicolumn{1}{c}{} &
\multicolumn{1}{c}{s} &
\multicolumn{1}{c}{10$^{-3}$ cps} &
\multicolumn{1}{c}{$\Gamma_{\rm X}$} &
\multicolumn{1}{c}{10$^{-13}$ erg cm$^{-2}$ s$^{-1}$} \\
\hline
04/04/2012 & 1948 & 4.3 $\pm$ 0.5 & $1.6 \pm 0.3$ & $2.7 \pm 0.9$ \\
18/04/2012 & 2630 & 2.5 $\pm$ 0.3 & $2.0 \pm 0.4$ & $1.2 \pm 0.3$ \\
15/11/2012 & 3679 & 2.4 $\pm$ 0.3 & $1.5 \pm 0.3$ & $1.4 \pm 0.2$ \\
\hline
\end{tabular}
\end{center}
\tablenotetext{a}{Observed flux}
\label{XRT}
\end{table*}

\section{Swift Data: Analysis and Results}
\label{SwiftData}

The {\em Swift} satellite \citep{gehrels04} performed three observations
of TXS\,0536$+$145 in 2012. The first two were carried out a few days after
the high $\gamma$-ray activity detected by {\em Fermi}-LAT at the end of
March. The last observation took place a few months later, to check
the flux variability. The observations were
performed with all three on board instruments: the X-ray Telescope
\citep[XRT;][0.2--10.0 keV]{burrows05}, the Ultraviolet/Optical Telescope
\citep[UVOT;][170--600 nm]{roming05} and the Burst Alert Telescope
\citep[BAT;][15--150 keV]{barthelmy05}.\\
The hard X-ray flux of this source turned out to be below the
sensitivity of the BAT 
instrument for such short exposures and therefore the data from this
instrument are not used.
It is worth mentioning that the source was not present in the {\em
  Swift} BAT 70-month hard X-ray
catalogue \citep{baumgartner13}.\\
The XRT data were processed with standard procedures (\texttt{xrtpipeline
  v0.12.6}), filtering, and screening criteria by using the
\texttt{HEAsoft} package 
(v6.12). The data were collected in photon counting mode in all the
observations. 
The source count rate
was low ($<$ 0.5 counts s$^{-1}$); thus pile-up correction was not
required. Source events were extracted from a circular region with a radius of
20 pixels (1 pixel $=$ 2.36\arcsec\ ), while background events were extracted
  from a circular region with radius of 50 pixels far away from the source
region and from other bright sources. Ancillary response files were 
generated with \texttt{xrtmkarf}, and
account for different extraction regions, vignetting and point-spread function
corrections. We used the spectral redistribution matrices v013 in the
calibration 
database maintained by HEASARC. Considering the small number of photons
collected ($<$200 counts) the spectra
were rebinned with a minimum of 1 count per bin and we used the Cash
statistic \citep{cash79}. We fitted the spectrum with an absorbed
power law
using the photoelectric absorption model \texttt{tbabs} \citep{wilms00}, with a
neutral hydrogen column density fixed
to its Galactic value \citep[2.79$\times$10$^{21}$ cm$^{-2}$;][]{kalberla05}.
The results are reported in Table~\ref{XRT}. The observations suggest
that TXS\,0536$+$145 was in a relatively bright state on 2012 April
  4, after the $\gamma$-ray flare, with a flux roughly a factor of 2
  higher than during April 18 and November 15. In
  the past, the 
  source was not detected by the {\em ROSAT} all-sky survey;
  thus this is the first detection in X-rays.\\
During the {\em Swift} pointing in 2012 April the UVOT instrument
observed TXS\,0536$+$145
in all its optical ($v$, $b$, and $u$) and UV ($w1$, $m2$, and $w2$) photometric
bands. On 2012 November 15 only observations with the V filter were
performed. We analyzed the data using the \texttt{uvotsource} task
included in the 
\texttt{HEAsoft} package. Source counts were extracted from a circular region of
5\arcsec\ radius centred on the source, while background
counts were derived from a circular region of 10\arcsec\ radius in the  source
neighbourhood. 
Due to
severe Galactic absorption and the short exposures the source was not
detected by UVOT in any of the filters. Upper limits are calculated
using the UVOT photometric
system when the analysis provided a detection significance $<$3
$\sigma$. As a reference on April 4 the lower limits on the magnitudes
are: $v$ $>$ 18.64, $b$
$>$ 19.82, $u$ $>$ 19.41, $w1$ $>$ 19.22, $m2$ $>$ 19.33, $w2$ $>$ 19.17. 

\begin{figure}
\begin{center}
\includegraphics{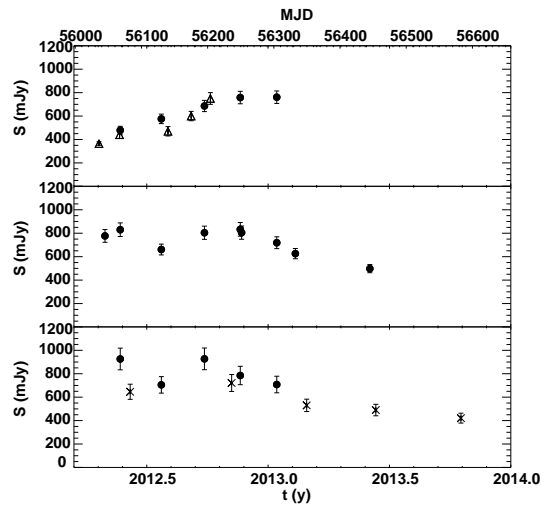}
\vspace{7cm}
\caption{Light curves at 8.4 GHz ({\it top}), 15 GHz ({\it middle}) 
and 24 GHz ({\it bottom}) of TXS\,0536$+$145. 
Filled circles are VLBA data, triangles are Medicina data at 8.4 GHz,
and crosses are EVN data at 22 GHz.}
\label{multi_light}
\end{center}
\end{figure}

\begin{figure}
\begin{center}
\includegraphics{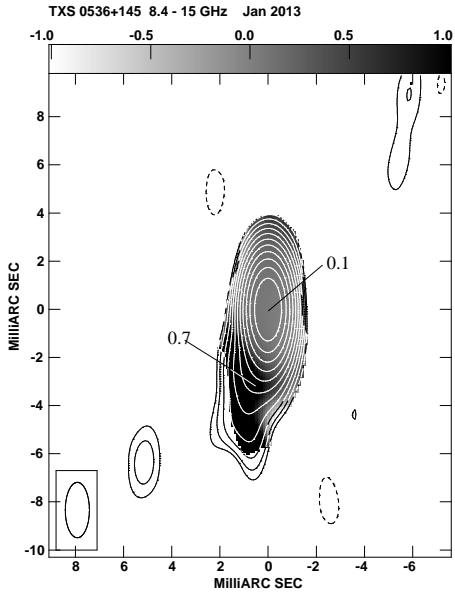}
\vspace{7.5cm}
\caption{Grey-scale spectral index image between 8.4 and 15 GHz
  superimposed on the low-resolution 8.4-GHz contours. The peak flux
  density is 720.9 mJy/beam. The first contour level is 0.5
  mJy/beam, which corresponds to three times the noise level measured
  on the image plane. Contour levels increase by a factor of 2. The 
restoring beam is plotted on the bottom left corner. The grey-scale is
shown by the wedge at the top of the spectral index image. }
\label{spix_image}
\end{center}
\end{figure}

\begin{figure}
\begin{center}
\includegraphics{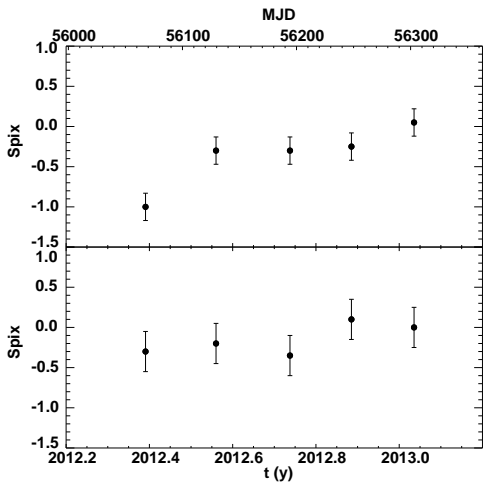}
\vspace{5.5cm}
\caption{Spectral index computed between 8.4 and 15 GHz ({\it upper
    panel}) and between 15 and 24 GHz ({\it lower panel}) for the five
VLBA observing epochs.}
\label{spix}
\end{center}
\end{figure}

\section{Results}

\subsection{Flux variability}

In 2012 TXS\,0536$+$145 showed strong variability throughout
the entire electromagnetic spectrum. The source was not present in the
first two {\it Fermi}-LAT catalogues \citep{abdo10a,nolan12} and no EGRET
$\gamma$-ray object was reported at the location of the
source. However, on 2012 March 22 the source underwent a major
$\gamma$-ray flare reaching an apparent isotropic luminosity of
6.6$\times$10$^{49}$ erg s$^{-1}$, which is the second most luminous
$\gamma$-ray flare from a blazar after 3C\,454.3 ($L_{\gamma}
\sim 2 \times 10^{50}$ erg s$^{-1}$, Abdo et al. 2011),
and is comparable
with those observed in other bright blazars, like PKS\,1510$-$089 ($L_{\gamma}
\sim 4 \times 10^{48}$ erg s$^{-1}$, Orienti et al. 2013),
PKS\,1830$-$211 ($L_{\gamma} \sim 3 \times 10^{49}$ erg s$^{-1}$, Abdo
et al., submitted), and 
PKS\,1622$-$297 ($L_{\gamma} \sim 4 \times 10^{48}$ erg s$^{-1}$, Mattox
et al. 1997). \\
Before this flare the source was first detected in $\gamma$
rays on 2012 January showing an enhancement of its high-energy activity, 
but without reaching a similar peak flux (Fig. \ref{LAT}).\\
Interestingly, a possible double hump is observed in the radio light
curves. From Fig. \ref{multi_light} it seems that the first peak in
the 15 GHz light curve occurs 
at the end of 2012 May, while the second peak is observed
at the end of 2012 September, although the time sampling of the
observations is not 
accurate enough to constrain more precisely the peak of the radio
outbursts. At 24 GHz the first peak is harder to constrain. 
At 8.4 GHz the variations are smoother and delayed by $\sim$2 
months with respect to what is observed at higher radio frequencies.
This agrees with the presence of opacity effects. 
In both high activity
states the increase of the $\gamma$-ray emission leads the radio
variability at 15 and 24 GHz with
a time lag of about 4--5 months.\\
The aim of the {\it Swift} observations was to unveil the counterpart
of the $\gamma$-ray flaring object, and not a detailed study of the
optical/X-ray variability.
{\it Swift} observations pointed out that in X-rays the source was in a
high activity state after the 2012 March flare, with a flux roughly
two times higher than what was observed in the subsequent observing
epochs. The high Galactic absorption precluded the detection in the UV
bands and in the optical U-band and B-band, 
and no information about the activity 
state at such wavelengths
could be obtained. The lack of detection in the V
band is likely due to the short exposure time of the {\it Swift}
observations. As a comparison, optical observations performed with
70-cm AZT-8 (CrAO,
Ukraine)\footnote{\url{http://lerga.crao.crimea.ua/Instr/azt8_en.html}} on
2012 
March 26, i.e. a few days after the peak of the $\gamma$-ray flare,
detected TXS\,0536$+$145 with a magnitude V=18.7$\pm$0.1 consistent
within the uncertainties, with the upper limit estimated by {\it
  Swift} observations.\\  

\begin{figure*}
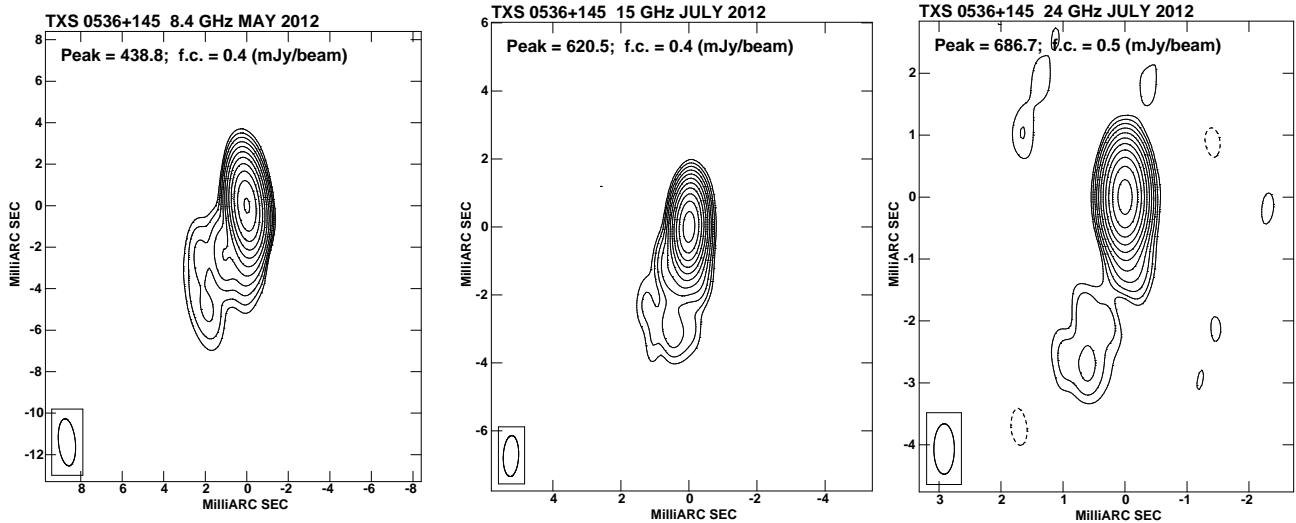

\begin{center}
\includegraphics{0536X_MAY12.PS}
\includegraphics{0536U_JUL12.PS}
\includegraphics{0536K_JUL12.PS}
\vspace{7cm}
\caption{VLBA images of TXS\,0536$+$145 at 8.4 GHz ({\it left}), 15 GHz
  ({\it center}), and 24 GHz ({\it right}). On each image, 
we provide the peak flux density in mJy/beam and the first contour
(f.c.) intensity in mJy/beam, which corresponds to three times the
off-source noise level. Contour levels increase by a factor of 2. 
The restoring beam is plotted in the bottom-left corner.}
\label{morphology}
\end{center}
\end{figure*}

\subsection{Spectral variability}

TXS\,0536$+$145 hardened its $\gamma$-ray 
spectrum during the flaring
episode. In 2012 March, i.e. the month of the highest activity, the photon
index was $\Gamma_{\gamma}$ =  2.05$\pm$0.08. This value is harder than the mean
photon index, $\Gamma_{\gamma}$ = 2.37$\pm$0.09 
derived between 2011 August and 2013 August. In addition, during the
period of highest activity, the spectral shape was better reproduced by a
log parabola model instead of a power law, indicating the presence of
a significant curvature in the $\gamma$-ray spectrum, which was not observed
during the average activity state. \\
No significant change in the photon index is observed in the X-rays.  \\
The radio spectral index analysis is a quite difficult task for VLBI, since it
is not possible to obtain a well matched {\it uv}-coverage at the various
frequencies, particularly at the short spacings. In addition, the
opacity of the source slightly changes the position of the core at the
various frequencies. With the aim of
studying changes in the spectral index distribution, in addition to the
full-resolution images, for each observing epoch 
we produced VLBA low-resolution images at each
frequency using the same {\it uv}-range (between 17 M$\lambda$ and 242
M$\lambda$), the same image
sampling, restoring beam, and natural grid weighting. An example of
spectral index image is presented in Fig. \ref{spix_image}.
Variations in the
spectral index computed considering the total flux density between 8.4
and 15 GHz, and between 15 and 24 GHz 
are reported in 
Fig. \ref{spix}. Between 8.4 and 15 GHz the spectrum was inverted
after the $\gamma$-ray 
flare, with a spectral index $\alpha_{8.4}^{15} = -1.0\pm0.2$,
then it flattened to a value $\alpha_{8.4}^{15} = 0.1\pm0.2$. At
the highest frequencies, the variation of the spectral shape is smoother
and the spectral index $\alpha_{15}^{24}$ ranges between $-0.3 \pm0.3$
and $0.1 \pm0.3$. 
Errors have been computed assuming the error
propagation theory.\\ 

\subsection{Radio morphology}

\begin{figure}
\begin{center}
\includegraphics{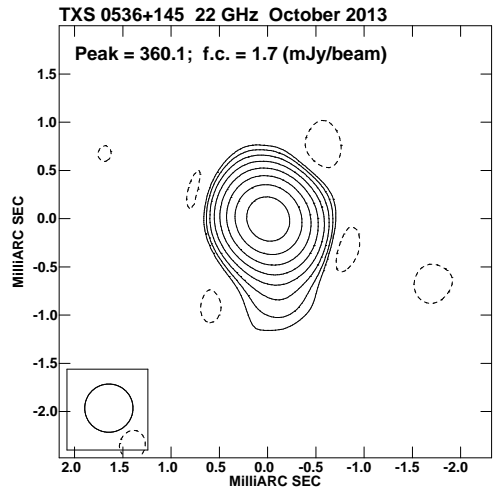}
\vspace{7cm}
\caption{EVN image of TXS\,0536$+$145 at 22 GHz. The image is restored
  with a circular Gaussian of FWHM = 0.5 mas, plotted on the
  bottom left corner. On the image
we provide the peak flux density in mJy/beam and the first contour
(f.c.) intensity in mJy/beam, which corresponds to three times the
off-source noise level. Contour levels increase by a factor of 2.}
\label{evn}
\end{center}
\end{figure}

When observed with parsec-scale resolution the radio source TXS\,0536$+$145 
shows a typical core-jet structure (Fig. \ref{morphology}). 
The radio emission is dominated by 
the compact bright component,
which accounts for about 90 per cent of the total flux density at 8.4 GHz, and about
95 per cent at 15 and 24 GHz.
The jet emerges from the main component 
 with a position angle of about 180$^{\circ}$, then at $\sim$1.5 mas
 (i.e. $\sim$12 pc) it slightly
 changes orientation to about 160$^{\circ}$ 
and extends to $\sim$6 mas (i.e. $\sim$48
 pc). Thanks to the higher angular resolution, the EVN image provides
 a deeper look into the jet base on sub-milliarcsecond scale
 (Fig. \ref{evn}), while lower sensitivity and worse sampling of the
 short spacing with respect to the VLBA prevent the detection of most
 of the extended emission from the jet. \\
No significant changes in the source structure are found
 by the analysis of the multi-epoch images.\\
Both the spectral and flux density variability arise
 from the core component as pinpointed by the multi-epoch
high angular resolution images. 
 The jet structure has a steep spectrum,
 $\alpha_{8.4}^{15}$=0.7$\pm$0.2, and does not show significant
 variability. \\
The similar flux density measured by the Medicina single-dish
telescope and the
VLBA at 8.4 GHz indicates that no significant emission extending
on kpc scale is present. \\

\section{Discussion}

The flux increase detected in $\gamma$-rays and at
lower energies is a strong indication that the flaring $\gamma$-ray
source is associated with the radio source TXS\,0536$+$145, becoming 
the $\gamma$-ray flaring blazar at the highest redshift observed so
far. The radio-to-$\gamma$-ray data collected by the multi-band
monitoring campaign triggered by the high energy flare 
allowed us to investigate the physical properties of this source and to
make a comparison with the characteristics of the high redshift 
population detected in $\gamma$-rays so far. \\

\subsection{The high activity state}

TXS\,0536$+$146 was in a $\gamma$-ray low
activity state during the first two years of {\it Fermi}-LAT 
observations. In 2012 the source entered in a high activity state:
it was detected on a monthly time scale for roughly the whole year
(Fig. \ref{LAT}, upper panel),
and on daily time scale in January and March during two major
outbursts (Fig. \ref{LAT}, bottom panel). In particular, during the 2012 March
flare the source reached an apparent isotropic $\gamma$-ray 
luminosity of about 6.6$\times
10^{49}$ erg s$^{-1}$, similar to the extreme luminosity reached by the
brightest FSRQ like 3C\,454.3 and PKS\,1510-089 \citep[e.g.][]{abdo11,mo13}.\\
In the radio band the light curves show a double enhancement which is
likely the counterpart at low energies of the two $\gamma$-ray
outbursts. The time sampling of the radio observations is not adequate 
for a proper determination of the flaring time. The time lag between
the $\gamma$-ray flares and the radio counterparts at 15 and 24 GHz 
is about 4 and 6
months in the observer frame for the first and second outburst
respectively, which means a delay of about
1 and 1.5 months in the source frame. This
value is consistent with the time-lag of 1.2 months found between
$\gamma$-ray and radio variability in a few samples of $\gamma$-ray
emitting blazars \citep{pushkarev10,fuhrmann14}.\\
The observing frequencies 8.4, 15, and 24 GHz 
correspond to 31, 55, and 88 GHz in the source frame,
where the opacity effects should be less severe. In addition,
observations at millimeter wavelengths of flaring blazars suggested that 
the onset of the radio
variability seems to lead the $\gamma$-ray flare, with a median time
delay in the source frame of about one month \citep{leon11}. The
lack of radio observations before the flaring episodes
precluded an accurate investigation of the radio light curves at 15 and 24
GHz before those episodes. By contrast, the availability
of archival data at 8.4 GHz allowed us to investigate the radio
behaviour from 2011 January to September. No evidence of flux density 
enhancement is found (Table \ref{histo}). This indicates that the time 
lag between the
beginning of the radio flare and the first $\gamma$-ray peak is shorter
than 5 months, which corresponds to approximately 1.3 months in the
source frame. \\
Strong $\gamma$-ray flares are often found to be associated with the
emergence of new superluminal jet components which should be the
observational manifestation of the shock propagating along the jet
\citep{jorstad01,mo13,jorstad13}. No evidence of a superluminal
component is found in TXS\,0536$+$145 
by the comparison of the multi-epoch images at 22 GHz. This may be
related to the high redshift of the target. We set an upper limit on
the apparent separation velocity $v_{\rm app}$ using:

\begin{equation}
v_{\rm app} = \frac{D_{\rm L}}{(1 + z)} \frac{\theta}{t_{\rm obs}}
\label{beta}
\end{equation}

\noindent where $\theta$ is the angular resolution, $t_{\rm obs}$ is the
time interval spanned by the EVN observations, $D_{\rm L}$ 
is the luminosity distance, and $z$ is the
redshift. If in Equation \ref{beta} we assume $\theta = 0.5$ mas
(i.e. roughly the resolution of the EVN
observations) and $t_{\rm obs} = 500$ days which corresponds to the
time interval spanned by the EVN observations, we estimate that the jet
component should have $v_{\rm app} \geq 35c$ to be detected at the resolution of our multi-epoch radio data. Such
high apparent velocity is not commonly observed even in the most
extreme blazars \citep{lister13,lister09b,jorstad01}. \\

\subsection{SED Modeling}

\begin{figure}
\begin{center}
\includegraphics{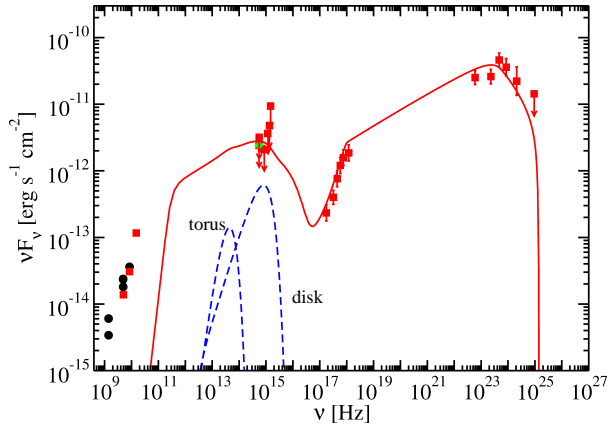}
\vspace{6cm}
\caption{SED data (squares) and model fit (solid curve) of
  TXS\,0536$+$145 in flaring 
  activity with the model components shown as dashed curves. Filled
  circles in the radio band are archival radio data, while the filled diamond in the optical regime refers to the detection in V-band on 2012 March 26 by the 70-cm AZT-8 telescope (see Section 5.1). }
\label{sed_fig}
\end{center}
\end{figure}

We modeled the SED of \txs\ with a combination of synchrotron,
synchrotron self-Compton (SSC), and external Compton (EC) non-thermal
emission.  We also included thermal emission by an accretion disc and
dust torus.  The modeling details can be found in \citet{finke08_SSC}
and \citet{dermer09_EC}. The synchrotron component considered in the
model is self-absorbed below 10$^{11}$ Hz and the radio data were not
fitted in the model.\\
The black hole mass for this object is not constrained, so we assumed its mass
to be $M_{BH}=10^9$M$_\odot$. Unfortunately, unlike the case of the
  high-redshift blazar S5\,0014$+$813 \citep{ghisellini09a} we are
  not able to estimate the accretion disc parameters from the optical
  data, since the optical data for TXS\,0536$+$145 are
upper limits with the exception of a single detection in the V-band (see Section 5.1). With these data we can only put an upper limit on the disc luminosity of about 
$10^{47}$\ erg s$^{-1}$, slightly below the Eddington limit for a
10$^{9}$ M$_\odot$ black hole, and we modelled the source in a
sub-Eddington accretion regime.  
We used a variability timescale of 1 day to
constrain the size of the emitting region, consistent with the
source's $\gamma$-ray light curve.  The results of the modeling can be
found in Table \ref{table_fit} and Figure \ref{sed_fig} \citep[see
][for a description of the model parameters]{dermer09_EC}. The model
has the magnetic field in equipartition, which means that the energy
stored in relativistic particles is approximately equal to the
magnetic energy.   
The EC seed photon source
was chosen to be those from the dust torus, 
and this external radiation field was
treated as an isotropic, monochromatic photon source. The dust torus
parameters were chosen to be consistent with the relation between
inner radius, disc luminosity, and dust temperature from
\citet{nenkova08}.  Due to the rather poor optical coverage, the model
parameters are not well constrained.

\begin{table*}
\footnotesize
\begin{center}
\caption{Model parameters.}
\label{table_fit}
\begin{tabular}{lcc}
\hline
Parameter & Symbol & Value  \\
\hline
Redshift & 	$z$	& 2.69	\\
Bulk Lorentz Factor & $\Gamma$	& 30	 \\
Doppler factor & $\delta_{\rm D}$	& 30	 \\
Magnetic Field [G]& $B$         & 1.2    \\
Variability Timescale [s]& $t_{\rm v}$       & 8.64$\times$$10^4$  \\
Comoving radius of blob [cm]& $R^{\prime}_{\rm b}$ & 2.1$\times$10$^{16}$  \\
\hline
Low-Energy Electron Spectral Index & $p_{\rm 1}$       & 2.5  \\
High-Energy Electron Spectral Index  & $p_{\rm 2}$       & 3.5  \\
Minimum Electron Lorentz Factor & $\gamma^{\prime}_{\rm min}$  & $3.0$  \\
Break Electron Lorentz Factor & $\gamma^{\prime}_{\rm brk}$ & $2.6\times10^3$  \\
Maximum Electron Lorentz Factor & $\gamma^{\prime}_{\rm max}$  & $2.0\times10^4$  \\
\hline
Black hole Mass [M$_\odot$] & $M_{\rm BH}$ & $1.0\times10^9$  \\
Disc luminosity [$\erg\ \s^{-1}$] & $L_{\rm disc}$ & $3.9\times10^{46}$ \\
Inner disc radius [R$_{\rm g}$] & $R_{\rm in}$ & $6.0$  \\
Seed photon source energy density [$\erg\ \cm^{-3}$] & $u_{\rm seed}$ & $1.3\times10^{-3}$  \\
Seed photon source photon energy & $\e_{\rm seed}$ & $1.0\times10^{-6}$  \\
Dust Torus luminosity [$\erg\ \s^{-1}$] & $L_{\rm dust}$ & $1.1\times10^{46}$  \\
Dust Torus radius [cm] & $R_{\rm dust}$ & $1.0\times10^{19}$  \\
Dust temperature [K] & $T_{\rm dust}$ & $2000$ \\
\hline
Jet Power in Magnetic Field [$\erg\ \s^{-1}$] & $P_{\rm j,B}$ & $4.3\times10^{45}$  \\
Jet Power in Electrons [$\erg\ \s^{-1}$] & $P_{\rm j,e}$ & $9.8\times10^{45}$  \\
\hline
\end{tabular}
\end{center}
\end{table*}

\subsection{The high-redshift $\gamma$-ray population}

\begin{figure}
\begin{center}
\includegraphics{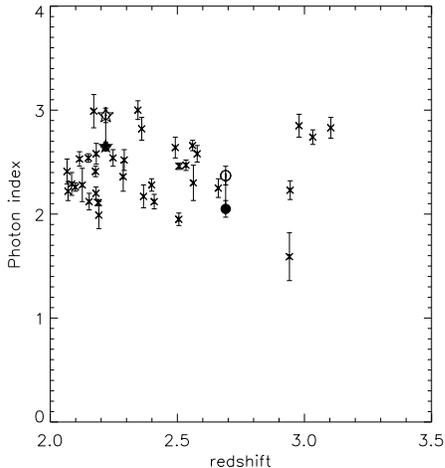}
\vspace{6.3cm}
\caption{Photon index versus redshift. Crosses are the
high redshift sources from the 2LAC with $z > 2$ 
\citep{ackermann11}, the empty and
filled circles are TXS\,0536$+$145 during the average activity
between 2011 August and  2013 August,  
and during the flare (2012 March), respectively. Empty and filled
stars refer to the high
redshift flaring object 4C\,$+$71.07 during the low activity state 
and the high activity state, respectively \citep{akyuz13}. } 
\label{phot-z}
\end{center}
\end{figure}

\begin{figure}
\begin{center}
\includegraphics{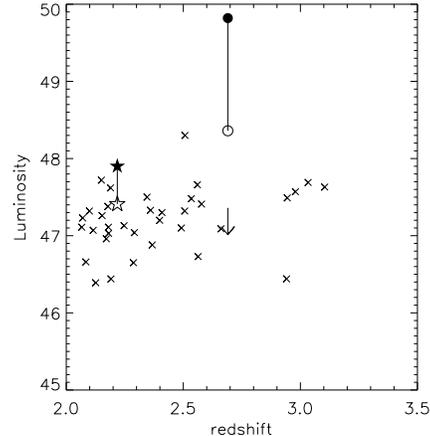}
\vspace{6.3cm}
\caption{Luminosity versus redshift. Crosses are the
high redshift sources from the 2LAC with $z > 2$ 
\citep{ackermann11}, the empty and
filled circles are TXS\,0536$+$145 during the average activity
between 2011 August and  2013 August,  
and during the flare (2012 March), respectively. The arrow is the
luminosity upper limit for TXS\,0536$+$145 computed during the low activity
state.
Empty and filled stars refer to the high
redshift flaring object 4C\,$+$71.07 during the low activity state and
during the flare, respectively \citep{akyuz13}.  } 
\label{lum-z}
\end{center}
\end{figure}

\begin{figure}
\begin{center}
\includegraphics{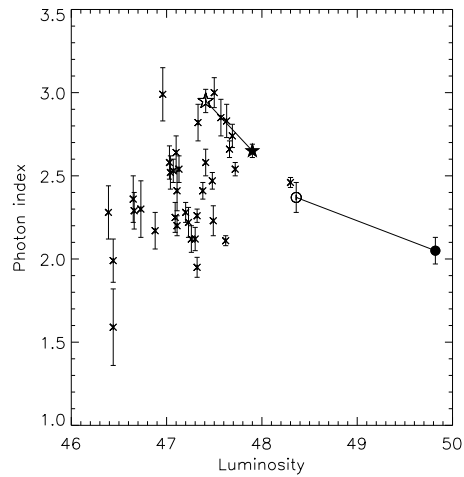}
\vspace{6.3cm}
\caption{Photon index versus luminosity. Crosses are the
high redshift sources from the 2LAC with $z > 2$ 
\citep{ackermann11}, the empty and
filled circles are TXS\,0536$+$145 during the average activity
between 2011 August and  2013 August,  
and during the flare (2012 March), respectively. 
Empty and filled stars refer to the high
redshift flaring object 4C\,$+$71.07 during the low activity state and the
flaring state, respectively \citep{akyuz13}.  }
\label{lum-gamma}
\end{center}
\end{figure}

To have a full characterization of the physical properties of
TXS\,0536$+$145, we compare its observational characteristics with those
shown by 
the population of high redshift ($z > 2$) $\gamma$-ray sources. To
this purpose we selected the $\gamma$-ray sources from the 2LAC that 
are associated at high confidence with a blazar at redshift $z >2$
\citep{ackermann11}. Sources with multiple associations
were discarded. The final sample of high-redshift $\gamma$-ray sources
consists of 35 FSRQ.\\
Fig. \ref{phot-z} shows the photon index versus the redshift for the
high-redshift sample. The photon index ranges between
1.6 and 3.0, with a distribution mean value
$\Gamma_{\gamma}=$2.4$\pm$0.3, which is
in agreement with the mean value $\Gamma_{\gamma}=$2.42$\pm$0.17
derived for the whole 
FSRQ population from 2LAC \citep{ackermann11}. The photon index of
TXS\,0536$+$145 is $\Gamma_{\gamma} \sim 2.4$, in excellent agreement with the
mean value, and it gets harder during the flare ($\Gamma_{\gamma} \sim
2.0$). \\
Among the high-redshift $\gamma$-ray sources, changes of the
photon index during different activity states were investigated for
the source 4C\,$+$71.07 (alias S5\,0836$+$71) by \citet{akyuz13}. This
source has a redshift $z=2.218$ and its photon index varies
between 2.95 and 2.65 during the low activity and flaring states,
indicating a
softer spectrum than that observed in TXS\,0536$+$145. It is worth
noting that during the flaring state the spectrum of both sources
deviates from a power law, showing a significant curvature and a
log-parabola shape.\\
New $\gamma$-ray flaring objects at high redshift are 
not commonly detected. So far 10 sources with $z > 2$ have been
observed during a $\gamma$-ray flare, and only TXS\,0536$+$145
is not part of the 2LAC. \\
Fig. \ref{lum-z} plots the $\gamma$-ray luminosity (E $>$ 100 MeV)
versus redshift, for high-redshift $\gamma$-ray sources in the 2LAC. The
$\gamma$-ray luminosity, $L_{\gamma}$ is computed following \citet{ghisellini09b}:

\begin{equation} 
L_{\gamma} = 4 \pi D_{\rm L}^2 \frac{S_{\gamma}}{(1+z)^{2-\Gamma_{\gamma}}}
\label{eq_lum}
\end{equation}

\noindent where $S_{\gamma}$ is the energy flux between 100 MeV and
100 GeV, and $\Gamma_{\gamma}$ is the
photon index. For the 2LAC sources, we calculated the $\gamma$-ray
luminosity considering in Eq. \ref{eq_lum} the energy flux
reported in the 2FGL \citep{nolan12}. The $\gamma$-ray luminosity
ranges between 2.5$\times 10^{46}$ erg s$^{-1}$ and 2$\times 10^{48}$
erg s$^{-1}$, with a median value $L_{\gamma}$ $\sim$1.6$\times$10$^{47}$ erg s$^{-1}$. As was already pointed out in \citet{ackermann11},
high-redshift objects tend to be the most luminous ones due to the
observational sensitivity limitation. In Fig. \ref{lum-z} we report
the luminosity of both TXS\,0536$+$145 and 4C\,$+$71.07 during the flare
and the average state. 
In addition, for TXS\,0536$+$145 we report the
upper limit computed by considering the first two years of {\it
  Fermi}-LAT observations, i.e. when the source was not detected. The
upper limit, $L_{\gamma} <2.3 \times 10^{47}$ erg s$^{-1}$, is
consistent with the luminosity values of the other high-redshift
blazars. \\
During the flare, TXS\,0536$+$145 increased its
$\gamma$-ray luminosity more than an order of magnitude with
respect to the average luminosity, reaching
$L_{\gamma} = 6.6 \times 10^{49}$ erg s$^{-1}$. Such a huge increase
of the $\gamma$-ray luminosity was observed from 3C\,454.3 during the
exceptional flare that occurred in 2010 November. The peak luminosity 
$L_{\gamma} \sim$2$\times$10$^{50}$ erg s$^{-1}$ was more than an order of
magnitude higher than the average luminosity $L_{\gamma} \sim 5\times
10^{48}$ erg s$^{-1}$ from the 2FGL \citep{nolan12}. \\
To investigate a change in the spectrum during the
high activity period, in Fig. \ref{lum-gamma} we plot the photon index
versus the $\gamma$-ray luminosity. It is clear that both
TXS\,0536$+$145 and 4C\,$+$71.07 show a ``harder when brighter'' effect,
in agreement with what was found for other variable blazars
\citep{abdo10c}. 
The hardening of the spectrum during a flare from a FSRQ 
should allow us to detect photons at higher energies than those
observed during an average state.
Considering the EBL model discussed in \citet{finke10}, at the
redshift of TXS\,0536+145 the optical depth should be $\tau \sim 1$
for 40 GeV photons.
The
maximum photon energy observed from TXS\,0536$+$145 is 11.2 GeV and is
consistent with the current EBL models. \\

\section{Conclusions}

In this paper we presented the results of the multi-wavelength campaign
for the flaring blazar TXS\,0536$+$145 carried out after two flaring
$\gamma$-ray events in 2012 January and March. 
This source was not part of the 1FGL
and 2FGL catalogues, indicating its low $\gamma$-ray activity state during the
first two years of {\it Fermi}-LAT observations.
During the brightest $\gamma$-ray flare the source
reached an apparent isotropic $\gamma$-ray luminosity of
6.6$\times$10$^{49}$ erg s$^{-1}$, comparable to the values achieved
by the most luminous and variable blazars. This episode triggered a
monitoring campaign from radio to X-rays performed by
VLBA, EVN, Medicina, and {\it Swift}. The spatial association
between the $\gamma$-ray source and the low energy counterpart,
together with the detection of multi-wavelength variability allowed us
to firmly identify the $\gamma$-ray object with TXS\,0536$+$145,
becoming the highest redshift $\gamma$-ray flaring source detected so far. \\
The analysis of the $\gamma$-ray {\it Fermi}-LAT light curve pointed out that
the source was first detected in a high activity state in 2012
January, without reaching the peak luminosity observed during the
following flare. A similar variability is observed in the radio band,
where a possible double hump is seen in the light curves at 15 and 24
GHz. However, the variability at the two extremes of the electromagnetic
spectrum seems delayed and the $\gamma$-ray leads the radio with a
time lag of about 4-6 months, which corresponds to about 1.5 months in
the source frame. \\
In the VLBI radio images this source displays a core-jet structure on the
pc scale. Both the flux density and spectral variability are
from the core component, while the jet does not show any
significant change. The delay between the $\gamma$-ray and radio
variability and its location in the unresolved central radio component
indicates that the $\gamma$-ray flare is produced in the innermost region of the
AGN close to the supermassive black hole, 
where the opacity severely affects the radio
emission. As the relativistic plasma expands, its emission becomes
optically thin at long wavelengths and the variability is observed in
the radio band.\\
No evidence of a newly ejected superluminal
component is found during the 16 months spanned by the VLBI monitoring
campaign. \\
The broad-band SED is well fitted by a synchrotron/external Compton model
where the seed photons may be those from the dust torus.
In the $\gamma$-rays the spectrum becomes harder
during the flare,
and deviates from a simple power law, showing a curvature that was not
observed in the average activity state. Despite the harder spectrum,
no significant emission above 10 GeV is observed.
The EBL should cause an energy-dependent suppression of the
  $\gamma$-ray flux from blazars that increases for larger
  redshifts. Such effect was observed in a sample of BL Lac object
  with redshift up to $z \sim 1.6$ \citep{ackermann12}. At the
  redshift of TXS\,0536$+$145 the flux attenuation 
should be observable below 10 GeV, leading to a
curvature of the spectrum. A significant spectral curvature was
observed during the 2012 March $\gamma$-ray flare, but the statistics are not
enough for testing if it is related to the EBL attenuation,
to the Klein-Nishina effect, or if it is intrinsic to the spectrum of
the source. The improved sensitivity of the LAT at a few GeV with Pass
8 data will be important for characterizing in more detail the
$\gamma$-ray spectrum of this high-redshift blazar.\\

\section*{Acknowledgments}
We thank D. Blinov for providing the 70-cm AZT-8 optical data. We
are grateful to D.J. Thompson and S. Digel for carefully reading the manuscript.
We thank the anonymous referee for useful suggestions.
The VLBA is operated by the US National Radio Astronomy Observatory
which is a facility of the National Science Foundation operated under
a cooperative agreement by Associated Universities, Inc.  
The European VLBI Network is a joint facility of European, Chinese,
South African and other radio astronomy institutes funded by their
national research councils. The research leading to these results has
received funding from the European Commission Seventh Framework
Programme (FP/2007-2013) under grant agreement No. 283393 (RadioNet3). 

The {\it Fermi} LAT Collaboration acknowledges generous ongoing support
from a number of agencies and institutes that have supported both the
development and the operation of the LAT as well as scientific data analysis.
These include the National Aeronautics and Space Administration and the
Department of Energy in the United States, the Commissariat \`a
l'Energie Atomique 
and the Centre National de la Recherche Scientifique / Institut
National de Physique 
Nucl\'eaire et de Physique des Particules in France, the Agenzia
Spaziale Italiana 
and the Istituto Nazionale di Fisica Nucleare in Italy, the Ministry
of Education, 
Culture, Sports, Science and Technology (MEXT), High Energy Accelerator Research
Organization (KEK) and Japan Aerospace Exploration Agency (JAXA) in Japan, and
the K.~A.~Wallenberg Foundation, the Swedish Research Council and the
Swedish National Space Board in Sweden. Additional support for science analysis during the operations phase is gratefully
acknowledged from the Istituto Nazionale di Astrofisica in Italy and the Centre National d'\'Etudes Spatiales in France.\\
This research has made use of the
data from the MOJAVE database that is maintained by the MOJAVE team
(Lister et al. 2009, AJ, 137, 3718). Part of the research is based on
observations with the Medicina telescope operated by INAF - Istituto
di Radioastronomia. We acknowledge the Enhancement Single-Dish Control
System (ESCS) Development Team at the Medicina telescope.  \\
This research has made use of the NASA/IPAC
Extragalactic Database NED which is operated by the JPL, Californian
Institute of Technology, under contract with the National Aeronautics
and Space Administration.

\end{document}